\documentclass[12pt]{iopart} 

\usepackage{color,amssymb, graphicx, amsfonts ,  setspace}
\usepackage{setstack}
\usepackage{enumerate}
\usepackage{myamsmath}
\usepackage{bm}

\newcommand{\nn}{\nonumber}
\newcommand{\me}[1]{\ensuremath{\mathrm{e}^{#1}}} %upright e
\newcommand{\dif}{\ensuremath{\,\mathrm{d}}} %upright d for dx
\newcommand{\av}[1]{\ensuremath{\left\langle #1 \right\rangle}}
\newcommand{\dro}[2]{\ensuremath{\frac{\dif{#1}}{\dif {#2}}}}

\begin{document}

\title{Upper critical field as a probe for multiband superconductivity in bulk and interfacial STO}
\author{J. M. Edge and A. V. Balatsky}
\address{Nordita, KTH Royal Institute of Technology and Stockholm University, Roslagstullsbacken 23
  106 91 Stockholm, Sweden,\\
Institute for Materials Science, Los Alamos National Laboratory, Los Alamos,
NM, 87545, USA}

\ead{edge@kth.se}

\date{\today}
\begin{abstract}
We investigate the temperature dependence of the upper critical field $H_{c 2} $ as a tool to probe the possible presence of multiband superconductivity  at the interface of LAO/STO. The behaviour of $H_{c 2} $ can clearly indicate two-band superconductivity through its nontrivial temperature dependence. For the  disorder scattering dominated two-dimensional LAO/STO interface we find a characteristic non-monotonic curvature of the $H_{c 2} (T) $. We also analyse the $H_{c 2} $ for multiband bulk STO and find similar behaviour.
\end{abstract}
\maketitle

\section{Introduction}
\label{sec:introduction}

Multiband superconductivity provides an intrinsically interesting extension of superconductivity. Shortly after the publication of BCS theory \cite{Bardeen1957},  an earliest idea of multiband superconductivity was proposed \cite{Suhl1959,Moskalenko1959}. It is characterised by having more than one band in which Cooper pairs form. Thus two different superconducting gaps may appear.
Apart from the theoretical interest, multiband superconductivity also has practical consequences. For example,
some of the highest temperature superconductors are multiband superconductors. These are magnesium diboride (MgB$_2$) with a transition temperature of 39 Kelvin \cite{Nagamatsu2001}, and the iron-based superconductors \cite{Kamihara2006,Kamihara2008}, with a maximal critical temperature of about 56 Kelvin \cite{Stewart2011}. Additionally, multiband superconductivity may lead to a higher upper critical magnetic field $H_{c 2} $ that is also attributable to the interplay between the two gaps \cite{Gurevich2003}.  Indeed, in the realm of technology applications, it has been speculated that due to these properties many future high magnetic field superconducting magnets, such as those found in MRI scanners, will be made of multiband superconductors \cite{Xi2008}.

Unambiguous detection of multiband superconductors requires advanced techniques.  Currently the main probes available are  scanning tunnelling spectroscopy \cite{Chuang2010}, heat transport \cite{Sologubenko2002,Seyfarth2005}, specific heat \cite{Zehetmayer2013} and the superfluid density \cite{Kim2002,Carrington2003}. Multiband superconductivity manifests itself through the occurrence of more than one quasiparticle coherence peak in tunneling spectroscopies \cite{Binnig1980}. However, short quasiparticle lifetimes may smear these peaks and thus make them unobservable. Heat transport may also be used to probe multiband superconductivity through its anomalous magnetic field dependence. A single band superconductor shows a strong suppression of heat transport all the way up to temperatures very close to the critical temperature. In contrast, in multiband superconductors one of the  gaps may be  disproportionately suppressed by a magnetic field, thus allowing that band to transport heat effectively \cite{Zehetmayer2013}. These techniques helped
determine that e.g. MgB$_2 $ \cite{Sologubenko2002} and PrOs$_4$Sb$_{12}$ \cite{Seyfarth2005} are multiband superconductors.

The recent discovery of superconductivity at the LaAlO$_3$/SrTiO$_3$ (LAO/STO) interface \cite{Reyren2007} has made the discussion of the nature of the superconducting state and possible multiband effects relevant \cite{Richter2013}. In this paper we wish to put forward the temperature dependence of the upper critical field as a probe for whether SrTiO$_3$ (STO) and particularly  the interface between LaAlO$_3$ (LAO) and STO are  single or multiband superconductors. The temperature dependence of the upper critical field may show characteristic behaviour inherent to multiband superconductivity and has been used previously to determine that iron-based superconductors are multiband superconductors \cite{Hunte2008}.

STO has long been a material of interest. It was the first oxide which was found to be superconducting \cite{Koonce1967}. Moreover, it was also the first material to show two-band superconductivity, through the presence of two quasiparticle coherence peaks \cite{Binnig1980}. STO can be tuned between single band and multiband superconductivity by changing the level of doping \cite{Binnig1980} and recently there have even been indications that, for certain doping levels, the material may be  a three-band superconductor\cite{Lin2013,Lin2014}. However, despite this evidence STO is still not unanimously accepted as a multiband superconductor \cite{Koshelev2003}.
Since 2004 attention has shifted to a metallic interface between LAO and STO \cite{Ohtomo2004}.  The system is remarkable since both LAO and undoped STO are insulators. Interest grew even further in 2007 when superconductivity was discovered at the interface \cite{Reyren2007}. One of the most pertinent questions now concerns the origin of the superconducting  state at the interface. 

One suggestion is that the metallic layer and thus the superconductivity is simply a consequence of surface doping at the interface \cite{Reyren2007}. However, in addition to the doping effects it was suggested that multiorbital effects \cite{Nakamura2013} and multiband effects are important \cite{Richter2013}  and in fact enable multiband superconductivity \cite{Fernandes2013}. The latter proposal, that the superconductivity is a direct descendant of superconductivity from the bulk STO, is supported by the fact that other interface layers apart from LAO also give rise to a metallic  and superconducting surface state of STO \cite{Perna2010,DiUccio2012}. Apart from the proposal of "descendant" superconductivity at the LAO/STO interface, the alternative   suggestions were made that the superconductivity at the surface is of an entirely different origin, resulting from a polar catastrophe and possibly spin orbit coupling that is a unique property of the interface and has no analog in bulk STO \cite{Nakagawa2006}. The ongoing debate  underscores the importance of unambiguous tests  that would clarify  the nature of the superconducting state. The investigation of  $H_{c2}(T)$ is one of these tests.

In this paper we  propose a direct test  of the hypothesis of two-band superconductivity in bulk STO and the LAO/STO interface. We consider  the perpendicular upper critical magnetic field in order to see if its behaviour can indicate whether the material is a single band or multiband superconductor. We concentrate on the upper critical magnetic field since it is a quantity readily accessible to experiments. Some other probes, like specific heat and heat transport,   are not practical for 
LAO/STO interfaces, thus making the temperature dependence of $H_{c2}$ one of the few available tools to further investigate superconducting states in these materials.  In doing so, we also aim to clarify the relationship between the superconductivity in the bulk and the interface system. 

The paper addresses both the case of bulk STO and LAO/STO interfaces. The possible regimes include four cases: clean and disordered (in the sense of the ratio of the coherence length to the mean free path) in bulk and interface STO.
We first investigate the dirty limit behaviour of the system. This is appropriate if the mean free path is shorter than the superconducting coherence length $\xi \approx 70 nm$ \cite{Reyren2007}; for interface systems this is likely to be the relevant situation. Depending on doping, it is also a realistic scenario for bulk STO, particularly at optimal doping \cite{Lin2013}.
Subsequently we address what is expected in a clean system. 
However, if there are two superconducting gaps, two coherence lengths and mean free paths are possible. 
In principle one could be in a regime where one band is dirty and the other is clean. This regime would require a complicated analysis and is outside the scope of this paper.  Our work expands and adds to   earlier work that focused on $H_{c 2} $, but only considered the clean limit \cite{Nakamura2013}.

In section~\ref{sec:band-structure-sto} we consider the band structure for STO and motivate our treatment of multiband superconductivity based on this band structure. In section~\ref{sec:details-calculation} we show how $H_{c2}(T) $ may be calculated for disordered multiband superconductors. Section~\ref{sec:results-h_c2-pres} presents the results for multiband superconductivity for coupling constants relevant to STO and also includes a more detailed investigation into the conditions under which $ H_{c2}(T) $ may be used to detect multiband superconductivity. In section~\ref{sec:clean_bulk_STO} we show the results for $H_{c2}(T)  $ in a clean system. Section~\ref{sec:lao_sto} first presents how the general calculation of section~\ref{sec:details-calculation} needs to be modified in order to consider the finite thickness of the superconducting layer at the LAO/STO interface. Subsequently the results for $H_{c2}(T)  $ are presented.

\section{The band structure of STO}
\label{sec:band-structure-sto}

Undoped STO has filled oxygen $p $ bands which are separated from the titanium $d $ bands by a large bandgap of 3eV \cite{Cardona1965}. Of these, the lowest result from the $ t_{2g}$ orbitals, $d_{xy},d_{yz} $ and $d_{xz} $, which get filled once the system is doped. The $ t_{2g}$ orbitals are split by the spin-orbit interaction and the crystal field. The highest energy band is situated approximately 30meV above a doublet of bands split by an amount of the order of 2meV \cite{VanderMarel2011}.

While this band structure may indicate that STO could form a one-, two- or three- band superconductor, we will investigate the distinction between single and two-band superconductivity only, as we wish to contrast single with multiband superconductivity. Furthermore, two of the bands are very close in energy and can thus easily couple together tightly and appear as a single band. 

Two important questions which will concern us are the couplings between the bands and the degree of anisotropy within each band. We will be primarily interested in the disordered limit, as described in section~\ref{sec:details-calculation}. Disorder scattering has the effect of averaging out Fermi surface anisotropies, such that one can effectively consider isotropic Fermi surfaces. In the clean limit the Fermi surfaces of STO are not perfectly isotropic, but for low degrees of doping we do not expect the anisotropies to be too great \cite{Lin2013}. 

Additionally, disorder scattering will introduce a coupling between the bands. We take this into account via the interband coupling constant in the self consistency equation (see section \ref{sec:details-calculation}). However, we do not expect this coupling to be very large and in particular, we expect it to be much smaller than any coupling within the bands. This is because the different $t_{2g} $ orbitals are orthogonal and have little spatial overlap. A coupling of the bands has to be able to effect an annihilation of a Cooper pair in one band and create it again in another. This process will be suppressed if the bands do not show great spatial overlap and is thus the justification for having a small interband coupling parameter, as described in section~\ref{sec:results-h_c2-pres}.

\section{Calculation of the upper critical field in the presence of disorder}
\label{sec:details-calculation}

At a quasi classical level the physics of a dirty superconductor can be described by the Usaldel equations \cite{Usadel1970}. These give an accurate description of the physics when disorder scattering is strong, such that anisotropies of the Fermi surface are averaged out. We solve the multiband Usadel equations \cite{Gurevich2003,Yerin2013} in the limit where the gaps $\Delta$ are very small.  This describes the region very close to the transition from superconductor to normal metal and the Usadel equations may be linearised, simplifying their solution. By solving the equations as a function of an applied magnetic field and temperature we thus obtain the temperature dependence of the upper critical magnetic field.

In our approach we closely follow the approach developed in Ref.\cite{Gurevich2003}.
We start with the linearised Usadel equations.
\begin{align}
  \label{eq:lin_usadel_eqn_wo_SOC}
  2\omega f_1 - D_1^{\alpha\beta} \Pi_\alpha \Pi_\beta f_1&=2\Delta_1\\
  2\omega f_2 - D_2^{\alpha\beta} \Pi_\alpha \Pi_\beta f_2&=2\Delta_2
\end{align}
$f_i $ , $i= 1,2$, is the Green's function of the system and in general depends on the momenta, position, and the Matsubara frequency $\omega = 2\pi T(2n +1)$. $D_i^{\alpha\beta} $ is the diffusivity tensor within a band. $\bm\Pi$ is defined as $\bm\Pi=\bm\nabla + 2\pi i \bm{A}/\phi_0$, $\phi_0$ is the flux quantum. By assuming the diffusivity tensor to be given by $D_m=\delta_{\alpha\beta}D_m $ and the vector potential to be given by $\bm A=H x \hat {\bm y}$, we can write these equations as
\begin{align}
 2\omega f_m- D_m
  \left(
    \nabla_x^2 + \nabla_y^2 + \nabla_z^2 + \frac{4\pi i H x}{\phi_0}\nabla_y \right.
  \nn \\
  \left.
    - \frac{4\pi^2 H^2 x^2}{\phi_0^2}
  \right) f_m
  = 2\Delta_m
  \; .
  \label{eq:expanded_usadel_eqn_wo_SOC}
\end{align}
Since this equation only depends on $x$, we now assume that $f_m$ is independent of $y$ and $z$ ($m \in \{1, 2\}$). Equation~\eqref{eq:expanded_usadel_eqn_wo_SOC} can now be solved for $\Delta_m$ and $f_m$ using the ansatz $f_m=h_m\Delta_m(x)$ and one obtains the solution
\begin{align}
  \label{eq:relation_f_and_Delta_wo_SOC}
  f_m(x,\omega)&=\frac{\Delta_m}{\omega+\pi H D_m/\phi_0}\\
  \Delta_m(x) &= \Delta_m' \me{- \pi H x^2 / \phi_0}
\end{align}
with $\Delta_m'$ being a constant.
The solutions for $f $ and $\Delta $ can be inserted into the gap equation for the two-band superconductor. This gives
\begin{align}
  \label{eq:4}
  \Delta_m&= 2 \pi T \sum_{\omega>0}^{\omega_D} \sum_{m'} \lambda_{m m'} f_{m'}(x, \omega)\\
  &=\sum_{m'} \lambda_{mm'} 2\pi T \sum_{\omega>0}^{\omega_D} \frac{\Delta_{m'}}{\omega + \pi H D_{m'}/\phi_0}\\
  &=\sum_{m'} \lambda_{mm'} \Delta_{m'}
  \left[
    \ln\frac{2 \gamma \omega_D}{\pi T} - U
    \left(
      \frac{H D_{m'}}{2\phi_0 T}
    \right)
  \right]
  .
\end{align}
$\omega_D$ is the Debye frequency and $\lambda_{mm'}$ the  superconducting coupling constants for the different bands.
In the last line we have used the equality
\begin{align}
  \label{eq:5}
  2\pi T \sum_{\omega>0}^{\omega_D} \frac{1}{\omega + X} = \ln\frac{2 \gamma \omega_D}{\pi T} - U
  \left(
    \frac{X}{2\pi T}
  \right)
\end{align}
with $U(x) = \psi(x+1/2) - \psi(1/2)$ and where $\psi$ is the di-gamma function. $\ln\gamma\approx 0.577$ is the Euler constant.
We can convert this into a 2x2 system of  equations for $ \Delta$ and divide out the factor $\me{- \pi H x^2 / \phi_0} $ and thereby replace $\Delta $ with $\Delta' $.
\begin{align}
  \label{eq:6}
  \underbrace{
  \begin{pmatrix}
    (l - U(h)) \lambda_{11} -1 & (l-U(\eta h))\lambda_{12}\\
    (l-U( h))\lambda_{21}  & (l - U(\eta h)) \lambda_{22} -1
  \end{pmatrix}
  }_{M_0}
  \begin{pmatrix}
     \Delta_1'\\  \Delta_2'
  \end{pmatrix}
  =0
\end{align}
Here $l=\ln\frac{2\omega_D \gamma}{\pi T}$, $h=\frac{H D_1}{2\phi_0 T}$ and $\eta=\frac{D_2}{D_1}$.

\begin{figure}
  \centering
  \includegraphics[width=8cm]{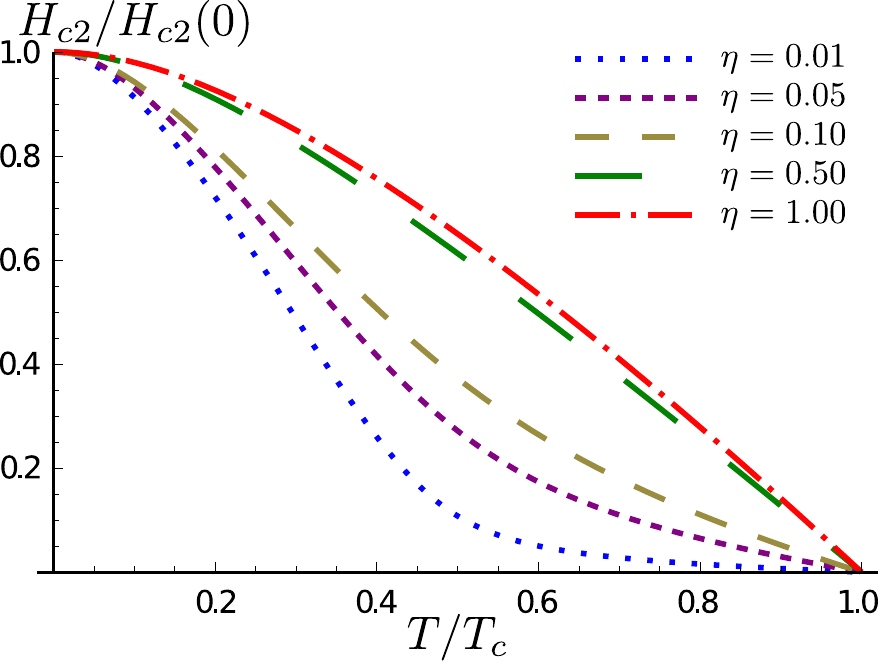}
  \caption{Temperature dependence of the upper critical field in the disordered limit for the set of coupling constants (\ref{enum:fern_pars}) \cite{Fernandes2013}. Different values of $\eta=D_2/D_1$ correspond to different ratios of the diffusivities.}
  \label{fig:Fs_params}
\end{figure}

Since these equations resulted from a linear expansion of the Usadel equations, they are valid for small, or infinitesimal $\Delta' $. Since $\Delta' $, and thus $\Delta$, is infinitesimal at $H = H_c $, these equations have a nontrivial solution at $H = H_c $. We thus need to find the solution to the equation $\det M_0 =0$.
After some manipulation one arrives at the expression
\begin{align}
  \label{eq:gures_cond_for_Hc2}
  a_0 (\ln t + U(h))(\ln t + U(\eta h)) + a_1 (\ln t + U( h))
  \nn \\
  +a_2 (\ln t + U(\eta h))  =0
  %.
\end{align}
with $t=\frac{T}{T_c}$. Here the equation for $T_c$ in a two-band superconductor has also been used in order to replace $\omega_D$ with $T_c$ (equation (22) in ref.\cite{Gurevich2003}). The coefficients $a_i$ depend of the coupling constants as follows
\begin{align}
  \label{eq:7}
  a_0&=\frac{2(\lambda_{11}\lambda_{22}-\lambda_{12}\lambda_{21})}{\lambda_0}\\
  a_1&=1+\frac{\lambda_{11}-\lambda_{22}}{\lambda_0}\\
  a_2&=1+\frac{\lambda_{22}-\lambda_{11}}{\lambda_0}\\
  \lambda_0&=\sqrt{\lambda_{11}^2 + \lambda_{22}^2+ 4 \lambda_{12}\lambda_{21}-2\lambda_{11}\lambda_{22}}
  .
\end{align}

\begin{figure}
  \centering
  \includegraphics[width=8cm]{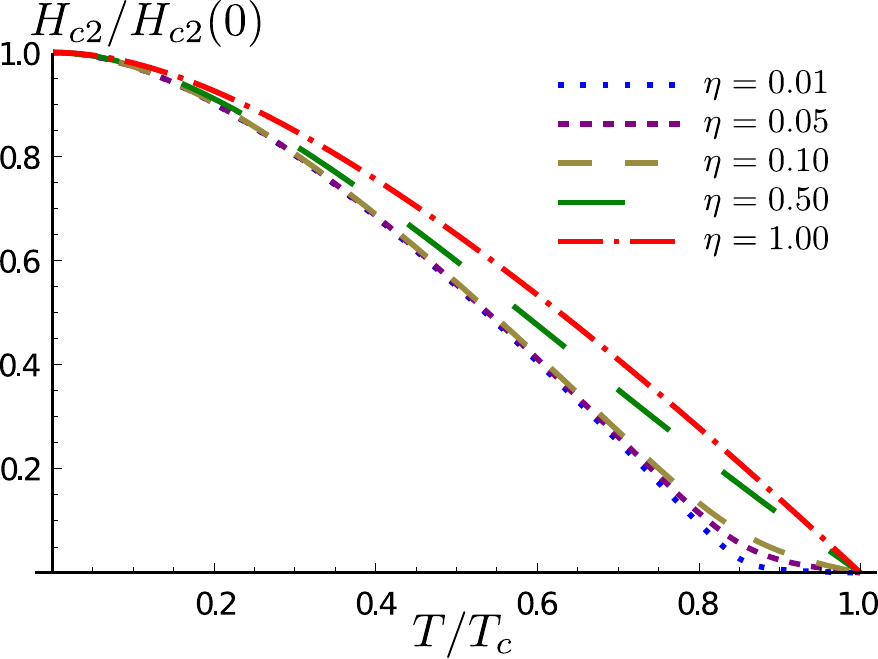}
  \caption{Same as figure~\ref{fig:Fs_params}, but for the set of  coupling constants (\ref{enum:BHs_pars}) \cite{Bussmann-Holder2010}.}
  \label{fig:BH_params}
\end{figure}

It is now relatively straightforward to solve numerically for the roots of equation~\eqref{eq:gures_cond_for_Hc2} as a function of $H_{c2} $ and $t=T/T_c$.

\section{Results for $H_{c2}$ in the presence of disorder}
\label{sec:results-h_c2-pres}

\subsection{Results for STO}
\label{sec:results-sto}

We now address the behaviour of $H_{c2}(T) $ as a function of the coupling constants and the diffusivity parameters. Our aim is to clarify under which circumstances $H_{c2}(T) $ may be used as a probe for multiband superconductivity. We first investigate the coupling constants applicable to STO.

There is no consensus for what the precise coupling constants for STO are. Two such sets are found in the literature:
\begin{enumerate}[(A)]
\item
  $\lambda_{11}=0.14, \lambda_{22}=0.13,\lambda_{12}=0.02$ \cite{Fernandes2013}
  \label{enum:fern_pars}
\item
   $\lambda_{11}=0.3, \lambda_{22}=0.1, \lambda_{12}=0.015$ \cite{Bussmann-Holder2010}
  \label{enum:BHs_pars}
\end{enumerate}
In figures~\ref{fig:Fs_params} and~\ref{fig:BH_params} we have plotted $H_{c2}(T) $ for the two sets of coupling constants (\ref{enum:fern_pars}) and (\ref{enum:BHs_pars}). Each plot contains the results for different ratios of the diffusivities in the two bands $\eta$.
If the diffusivities are the same in the two bands ($D_1=D_2$), the $ H_{c2}(T) $ curves are identical to those in single-band superconductors. Only once the diffusivities start to differ appreciably, do the $H_{c2}(T)  $ curves show a departure from single band behaviour. The characteristic two-band property of the $ H_{c2}(T) $ curves, and thus the indicator for the presence of two-band superconductivity, is a change in the curvature of the $H_{c2}(T)  $ curve, as can be seen most clearly in the blue dotted curve of figure~\ref{fig:Fs_params}. In the vicinity of $T_c $, $H_{c 2} $ initially grows very slowly, but at some temperature (here at $T\approx 0.5 T_c $) it starts growing dramatically until it saturates at $T = 0 $. In contrast, for single band superconductors or for equal diffusivities, $H_{c2}(T)  $ starts growing rapidly at $T_c $ and as $T\to0 $ the growth rate monotonically decreases (see red curve in figure~\ref{fig:Fs_params}).

While we have no precise calculation for the ratio of the diffusivity, at constant mean free time $\tau$ the diffusivities should be proportional to the square of the Fermi velocity, since $ D\sim l_{mfp}^2/\tau=\tau v_F^2$ ($l_{mfp}$ is the mean free path). The Fermi velocities in STO differ by about a factor of 3 or 4 between the two bands \cite{VanderMarel2011}. Assuming the mean free time  to be the same, we thus obtain a ratio of diffusivities of about 10, which is sufficient to observe the non-monotonic behaviour of the $ H_{c2}(T) $ curvature.

As we can clearly see from figures~\ref{fig:Fs_params} and~\ref{fig:BH_params}, the shape of the $H_{c2}(T)  $ curves depends strongly on the values of the coupling constants chosen. The set of coupling constants~(\ref{enum:fern_pars}) is much more favourable for the detection of multiband superconductivity than the set~(\ref{enum:BHs_pars}).

As we cannot be sure which set of coupling constants are precisely applicable for STO, we now turn to a broader investigation of the upper critical field for more general coupling constants.

\subsection{More general parameter values}
\label{sec:more-gener-param}

Here we explore in greater detail under which more general conditions two-band superconductivity can lead to a discernible modification of the $ H_{c2}(T)$ curve with respect to the single band behaviour. In exploring this behaviour we explicitly go beyond the values of the coupling constants expected for STO.
We concentrate on the physics of the bulk, as the physics of the interface is similar, as described in section~\ref{sec:lao_sto}.

\begin{figure}[tb]
  \centering
  \includegraphics[width=8cm]{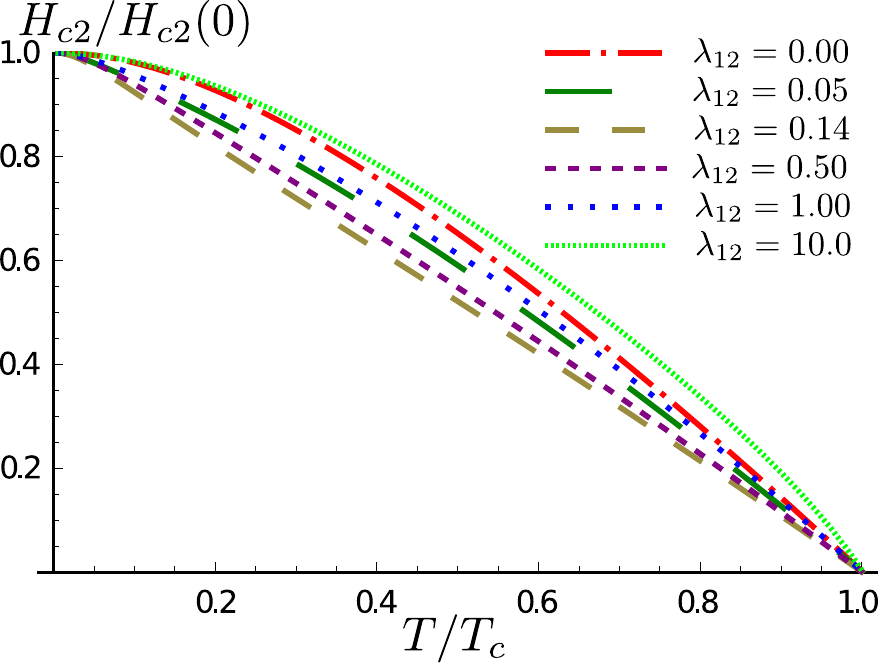}
  \caption{Upper critical field in the case where one of the intra-band coupling constants is zero. The parameters are given by $\lambda_{11}= 0.14, \lambda_{22}=0, \eta=0.1$.}
  \label{fig:l22=0_vary_l12}
\end{figure}

We first investigate the possibly simplest situation in which one of the coupling constants is zero, see figure~\ref{fig:l22=0_vary_l12}. We choose $\lambda_{22} =0$. In this case superconductivity only exists in the second band as a result of the induced superconductivity due to $\lambda_{12} $. For $ \lambda_{12}=0$ one obtains the single band $ H_{c2}(T)$ behaviour. Although there is a dependence of the curves on $ \lambda_{12}$, it is not very strong. Without access to the entire temperature range $0<\frac{T}{T_c}<1$ it would be difficult to conclude whether or not multiband superconductivity is present. The strongest departure from the single band behaviour of $H_{c2}(T) $ occurs at $\lambda_{12}\approx \lambda_{11} $. If $\lambda_{12}\gg\lambda_{11} $ the two bands are strongly locked to each other and thus the behaviour is similar to that for a single band system again. 

\begin{figure}[tb]
  \centering
   \includegraphics[width=8cm]{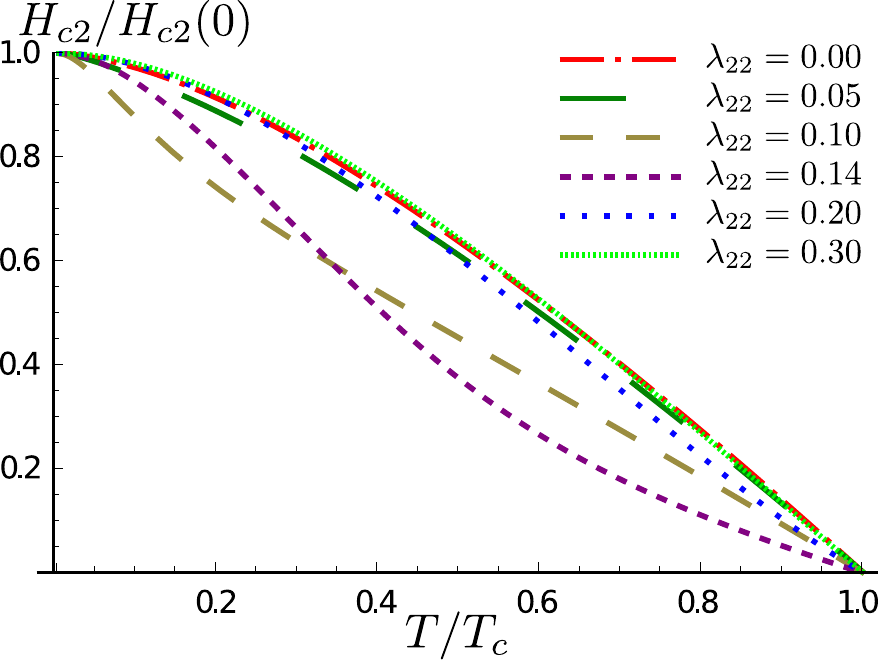}
  \caption{Upper critical field for the case in which a small interband coupling is chosen, and one of the intraband coupling constants is varied. The parameters are given by $\lambda_{11}=0.14, \lambda_{12}=0.02, \eta=0.1$. The strongest departure from single-band behaviour is observed when $\lambda_{11}\approx \lambda_{22}$.}
  \label{fig:l12_small_vary_l22}
\end{figure}

In figure~\ref{fig:l12_small_vary_l22} we fix $\lambda_{12} =0.02$, $\lambda_{11} =0.14$ and vary $\lambda_{22} $. We observe that the departure of the $ H_{c2}$ curve from single band behaviour is strongest when the coupling constants within the bands are roughly equal. If their difference is too great, one of the bands always dominates and the interplay of the two bands, which ultimately causes the non-monotonic curvature of $H_{c2}(T)$, cannot be observed.

\begin{figure}[tb]
  \centering
  \includegraphics[width=8cm]{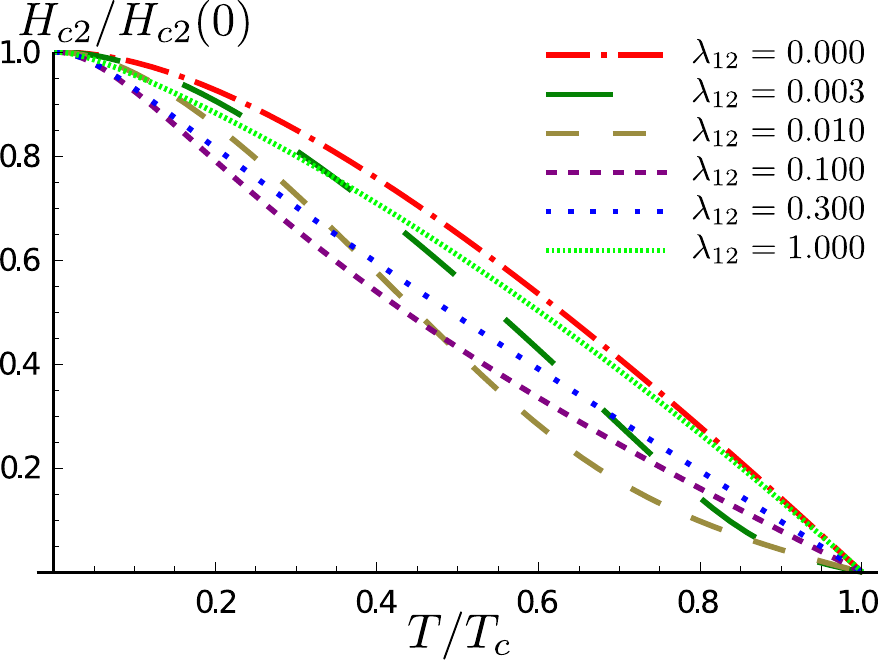}
  \caption{Upper critical field for the case $\lambda_{11}=\lambda_{22}=0.14 $ and $\eta=0.1$, in which now the inter-band coupling $\lambda_{12}$ is varied. The circumstances most favourable for the detection of multiband superconductivity ar when $\lambda_{12} $ is much smaller than $ \lambda_{11}$, but non-zero.}
  \label{fig:fix_l11=l22_vary_l12}
\end{figure}

In figure~\ref{fig:fix_l11=l22_vary_l12} we explore the behaviour of $H_{c2}(T) $ for different values of $\lambda_{12} $ in the case when $\lambda_{11}=\lambda_{22} $, the case most favourable for the detection of the signature of multiband superconductivity in $H_{c2}(T) $. If the coupling between the bands is absent, each band just shows single band behaviour and there is no signature of multiband superconductivity in the upper critical field. This is due to the fact that it is only the most dominant band, the one with the larger coupling constant, which determines $ H_{c2}$. 
As can be seen from figure~\ref{fig:fix_l11=l22_vary_l12}, the signature in the upper critical field can be best detected when $\lambda_{12} $ is significantly smaller than $\lambda_{11}=\lambda_{22} $, but non-zero. For $\lambda_{12}\approx\lambda_{11} $ a signature remains but requires access to a very large range of $ \frac T{T_c}$ for it to be detected.
Once $ \lambda_{12}\gg \lambda_{11}$ the bands are so strongly coupled that the system effectively behaves like a single band system.

From the above we may conclude that multiband superconductivity can be most easily detected through measurements of the upper critical field when the coupling constants within the two bands are approximately the same, the inter-band coupling constant is significantly smaller than the intra-band coupling constants, and the diffusivities in the two bands differ by at least a factor of 5.

We thus find that depending on what set of diffusivities and which of the two coupling constants are realised in real STO, two band superconductivity might be inferred from the shape of the $H_{c 2} (T) $ curve.  This observation can provide guidance for the search of multiband superconductivity in STO. On the other hand, a seemingly trivial behaviour of the  $H_{c 2} (T) $ curve does not imply that STO is a single band superconductor. 
It has been argued that unconventional $H_{c 2} $ behaviour could be expected even for a single band systems, as long as the single band is highly anisotropic \cite{Zehetmayer2013}. However, for STO this is not expected to be the case \cite{Lin2013,VanderMarel2011},  and an unconventional behaviour of $H_{c2}$ can be taken to be good evidence for multiband superconductivity.

\section{$H_{c2}$ for clean doped bulk STO}
\label{sec:clean_bulk_STO}

For completeness we also present the case of clean bulk superconducting STO. Away from optimal doping, bulk STO may enter a regime in which the mean free path is larger than the superconducting coherence length \cite{Lin2013}. In this regime a calculation for the clean system is more appropriate. We therefore briefly present the results obtained from the quasi-classical Eilenberger equations.
The critical field for a three-dimensional clean two-band superconductor is given by the solution of equation (76) in Ref.\cite{Kogan2012}
\begin{align}
  \label{eq:10}
  (\ln t)^2 - 2h_c(n_1 \alpha_{11} {\cal I}_1 + n_2 \alpha_{22} {\cal I}_2) \ln t \nn\\
  +4 h_c^2(n_1 \alpha_{11} + n_2 \alpha_{22} -1)  {\cal I}_1  {\cal I}_2=0\\
  {\cal I}_\beta=\int_0^\infty \dif s s \ln(\tanh(st)) \av{\mu_{c,\beta}\me{-\mu_{c,\beta} s^2 h_c}}_\beta
  .
\end{align}
$\av{\dots}_\beta$ is an average over the Fermi surface associated with the band $\beta \in \{1,2\}$ and $\mu_c=(v_x^2+v_y^2)/v_0$ with $v_0=(2E_F^2/(\pi^2\hbar^3 N_\beta))^{1/3}$. For isotropic bands $v_0=v_F$. $N_\beta$ is the density of states at the Fermi surface in band $\beta$. Since the bands are expected to be roughly isotropic, we will replace the average over $\mu_{c,\beta}$ with just a single (band dependent) value $\mu_{\beta}$. This we will vary, in order to explore the different types of behaviour.
$\alpha_{ii}$ are normalised coupling constants. They are normalised to the value of an effective coupling constant $\alpha_0$ whose value would determine the superconducting gap and hence $T_c$, if the system were a single band superconductor. $\alpha_0$ is thus given by \cite{Kogan2012}
\begin{align}
  \label{eq:11}
  \alpha_0&=
  \left(
    - \ln\frac{\pi \gamma T_c}{2\hbar \omega_D}
  \right)^{-1}
\end{align}
where $\ln \gamma$ is again the Euler constant and $\omega_D$ is the Debye frequency. $\alpha_{11}$ and $\alpha_{22}$ are accordingly given by
\begin{align}
  \label{eq:12}
  \alpha_{11}&=\lambda_{11}/\alpha_0\\
  \alpha_{22}&=\lambda_{22}/\alpha_0
  \; .
\end{align}

\begin{figure}[tb]
  \centering
  \includegraphics[width=8cm]{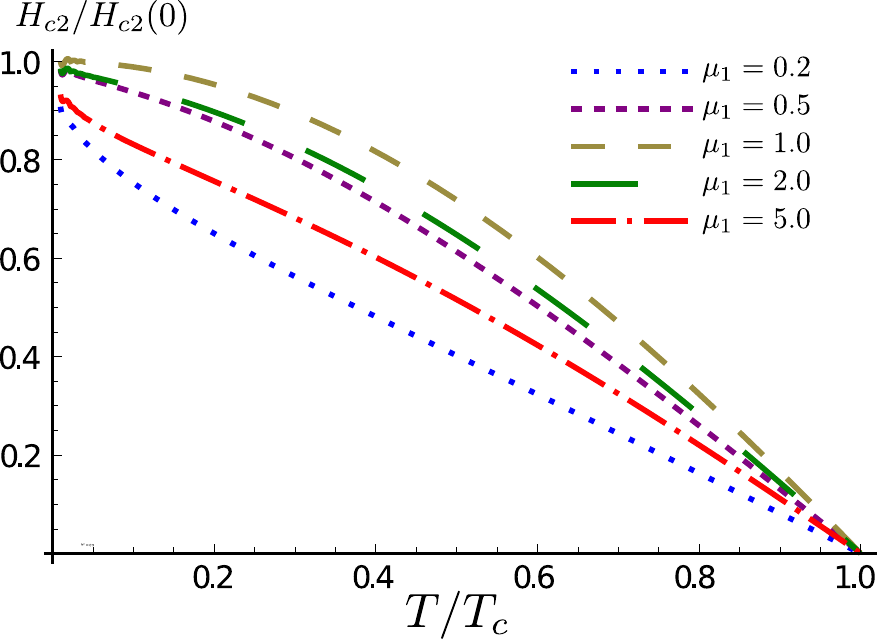}
  \caption{Temperature dependence of the upper critical field in the clean limit for the  coupling constants $\lambda_{11}=0.14, \lambda_{22}=0.13,\lambda_{12}=0.02$ \cite{Fernandes2013} (the same as in figure~\ref{fig:Fs_params}). We have set the parameter $\mu_2=1$ and vary the remaining parameter $\mu_1$.}
  \label{fig:clean_Fs_params}
\end{figure}

\begin{figure}[tb]
  \centering
  \includegraphics[width=8cm]{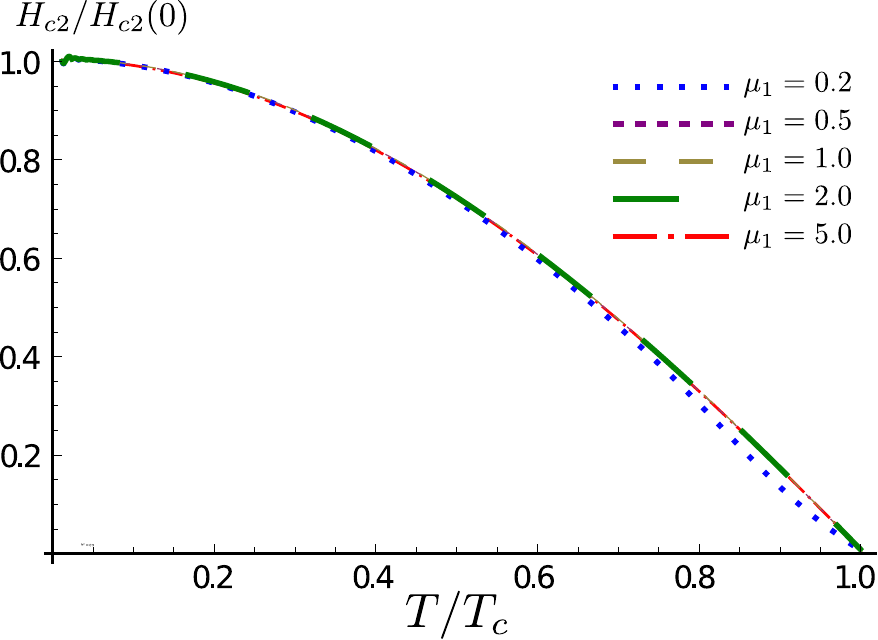}
  \caption{ Same as figure~\ref{fig:clean_Fs_params}, but for the coupling constants $\lambda_{11}=0.3, \lambda_{22}=0.1, \lambda_{12}=0.015$ \cite{Bussmann-Holder2010} (the coupling constants are the same as in figure~\ref{fig:BH_params}).}
  \label{fig:clean_BHs_params}
\end{figure}

For different values of the parameter $ \mu_{c,\beta}$ we have plotted the temperature dependence of $ H_{c2}$ in figures~\ref{fig:clean_Fs_params} and~\ref{fig:clean_BHs_params}. This is done again for two different values of the coupling constants found in the literature \cite{Fernandes2013,Bussmann-Holder2010}. We can see that these curves by and large do not give a clear indication of the presence of two-band superconductivity, at least for the temperature range which might be accessible to experiments. Therefore, it seems that the upper critical field can only be used to identify multiband superconductivity in STO in the dirty limit.

\section{$H_{c2}$ for the LAO/STO interface}
\label{sec:lao_sto}

The interface between LAO and STO is closer to the disordered limit than bulk doped STO. The mean free path in such a system has been estimated to be 25 nm\cite{BenShalom2009}  as opposed to approximately 60 nm for the bulk system at optimal doping \cite{Lin2013}. Therefore a calculation for the dirty system becomes necessary in this case, which complements the clean calculation that was performed previously \cite{Nakamura2013}. In the following we compute what is expected for a superconducting layer confined to a thickness $d $. From this we can then estimate the behaviour of the interface system under an applied magnetic field perpendicular to the interface.

We present a mean-field calculation of the upper critical field here. Although, strictly speaking, a BKT analysis of the interface would be more appropriate, the overall behaviour of the BKT transition will be determined by the mean-field value of the gaps.

At the interface we need to take account of two additional effects compared to the bulk: on the one hand, the electron gas and thus the superconductor is confined to a thickness $d $. On the other hand, it has been reported that due to the inversion symmetry breaking a Rashba spin-orbit coupling emerges at the interface \cite{Caviglia2010,BenShalom2010}. We will take the finite thickness of the layer into account by retaining the $\nabla_z^2 $ term in equation~(\ref{eq:expanded_usadel_eqn_wo_SOC}). In order to treat the effects of spin-orbit coupling, equation~(\ref{eq:expanded_usadel_eqn_wo_SOC}) needs to be generalised to a matrix equation with anomalous Green's function $\tilde f$.
Since the linearised Usadel equations do not couple the different bands directly, we may treat each band separately. In the following we will suppress the band index $m$ in order to simplify the notation.

\begin{figure}[tb]
  \centering
  \includegraphics[width=8cm]{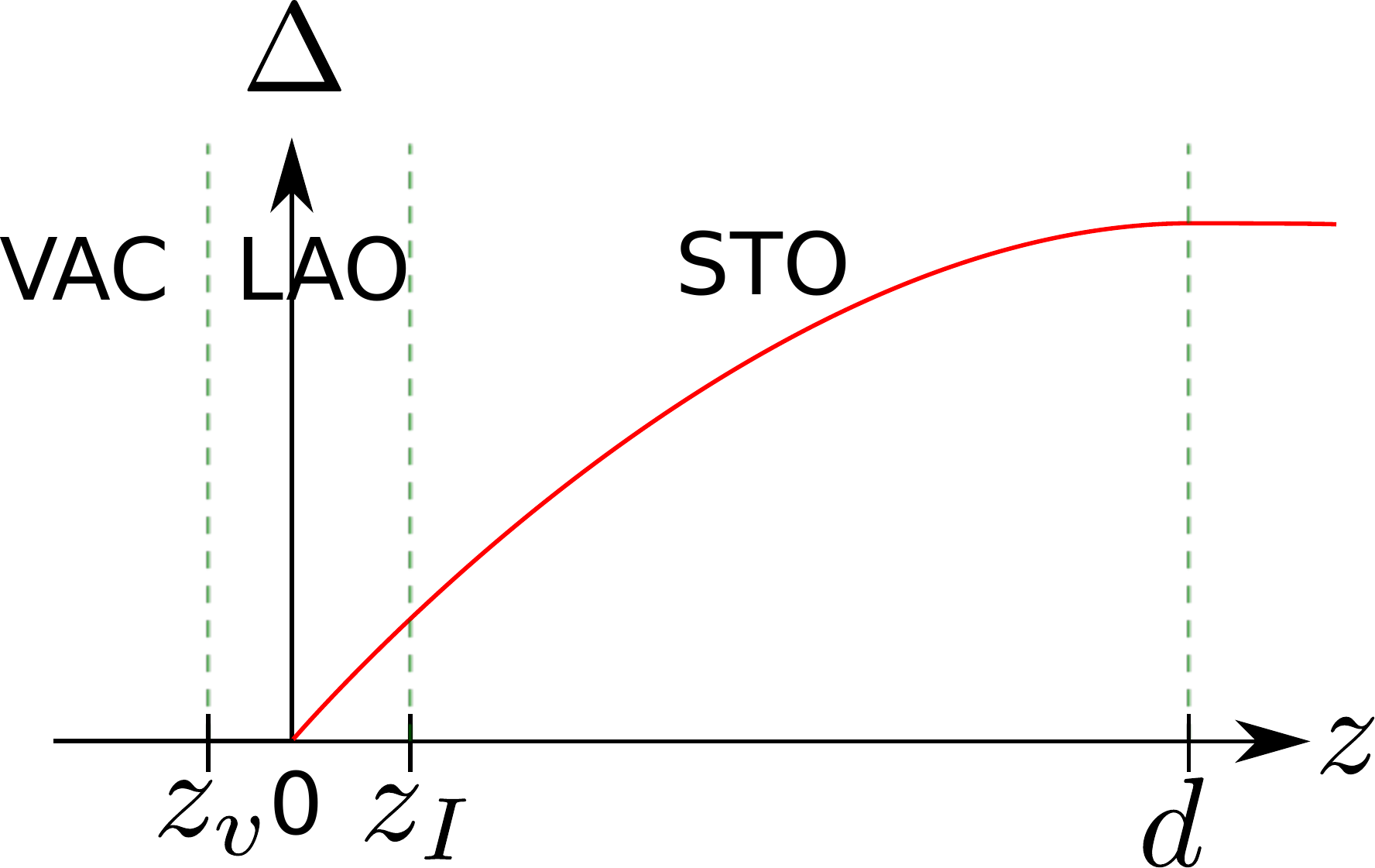}
  \caption{Schematic illustration of the geometry under consideration. The gap $\Delta$ needs to vanish somewhere at the LAO side of the interface layer, though since the LAO layer is much thiner than the width of the superconducting layer, it does not matter where exactly we specify $\Delta(z)=0$. $d$ is the width of the superconducting layer, $z_I$ the position of the interface between LAO and STO, and $z_v$ the position of the LAO-vacuum (or air) interface. At $z=d$ the superconductor is interfaced with a metal.}
  \label{fig:schematic_gap_at_interface}
\end{figure}

In the presence of Rashba spin orbit coupling, the operator $\bm\Pi$ becomes \cite{Bergeret2013}
\begin{align}
  \label{eq:def_of_tilde_Pi}
  \tilde {\bm \Pi}=
  \left(
    \nabla_x\sigma_0 + SO_x, \nabla_y \sigma_0  + \frac{2\pi i H x}{\phi_0} \sigma_0 + SO_y, \nabla_z\sigma_0  + SO_z
  \right)
\end{align}
The terms $SO_x $ and $SO_y $ are defined by their action $SO_x \tilde f=i\nu [\sigma_y, \tilde f]$, $SO_y \tilde f=-i\nu[\sigma_x, \tilde f] $. The strength of the spin-orbit interaction $\nu$ is related to the Rashba coupling term $\alpha$ by $\nu={\alpha m_e}/\hbar $ where $m_e$ is the mass of the electron.
$ \tilde f$ can be expanded into singlet $f_s$ and triplet $\bm f_t$ components
\begin{align}
  \tilde f &= i \sigma_y f_s + i\sigma_y \bm f_t \cdot\bm \sigma= i \sigma_y f_s + i\sigma_y (f^a, f^b, f^c)\cdot \bm{\sigma} \nn\\
  &
  =  i \sigma_y f_s + f^a \sigma_z + i f^b \sigma_0 - f^c \sigma_x
  .
  \label{eq:def_tilde_f}
\end{align}
Similarly, the superconducting gap $\Delta $ becomes a matrix $\tilde \Delta$, and can also be expanded as
\begin{align}
  \tilde \Delta &= i \sigma_y \Delta_s + i\sigma_y \bm \Delta_t \cdot\bm \sigma= i \sigma_y \Delta_s + i\sigma_y (\Delta^a, \Delta^b, \Delta^c)\cdot \bm{\sigma}\nn\\
  &=  i \sigma_y \Delta_s + \Delta^a \sigma_z + i \Delta^b \sigma_0 - \Delta^c \sigma_x
  \label{eq:delta_sing_trip}
\end{align}
In the absence of spin-orbit coupling only the dominant singlet component is relevant, so in section~\ref{sec:details-calculation}, $f $ and $\Delta$ could be treated as  scalars.
The Usadel equation for the single band (previously equation~(\ref{eq:expanded_usadel_eqn_wo_SOC})) then becomes
\begin{align}
  & 2\omega \tilde f 
  - D
  \left\{
     \nabla_x^2 \tilde f + \nabla_z^2\tilde f
  -4\nu\sigma_x \nabla_x f^a - 4\nu \sigma_z \nabla_x f^c
    - \frac{4\pi^2 H^2 x^2}{\phi_0^2}\tilde f 
  \right.
 \nn \\* & \qquad \qquad
  \left.
  + \frac{8\pi i \nu H x}{\phi_0}(i \sigma_z f_s - \sigma_y f^a)
  -4\nu^2(i\sigma_y f_s + 2\sigma_z f^a - \sigma_x f^c)
\right\}
\nn\\* & \qquad \qquad \qquad \qquad  \qquad \qquad \qquad \qquad 
= 2
\left(
  i\sigma_y \Delta_s + \Delta^a \sigma_z + i \sigma_0 \Delta^b - \sigma_x \Delta^c
\right)
  \label{eq:usadel_eqn_all_sing_trip_cmpts}
\end{align}
As we argue in appendix~\ref{sec:induc-tripl-superc}, for our purposes we may assume that the triplet components of $\tilde f $ and $ \tilde \Delta$ are zero. This considerably simplifies equation~(\ref{eq:usadel_eqn_all_sing_trip_cmpts}) and we therefore need to solve the equation
\begin{align}
    2\omega f_s - D
  \left\{
    \nabla_x^2 f_s - \frac{4\pi^2 H^2 x^2}{\phi_0^2} f_s - 4\nu^2 f_s
    + \nabla_z^2 f_s
  \right\} &= 2\Delta_s
  .
\end{align}
From now onwards we will suppress the index $s$. 
Compared with equation~(\ref{eq:expanded_usadel_eqn_wo_SOC}) (where $\nabla_y=\nabla_z=0$) the two new terms are $-4\nu^2 f$ and the $\nabla_z^2 f$ term. The term $-4\nu^2 f$ just results in a constant shift, represeting virtual processes of scattering to triplet components of the gap and back again to the singlet component.
In order to include the term $\nabla_z^2 f$ we modify the ansatz $f=h\Delta^x(x)$ from section~\ref{sec:details-calculation} to $f=h \Delta^x (x) \Delta^z (z) $ and specify boundary conditions for $\Delta^z(z)$.
The LAO forms a thin layer (for a typical 5 unit cells of LAO its thickness is 2nm \cite{Basletic2008}) of wide bandgap insulator material and borders air or vacuum. Since it is much thinner than the superconducting layer $d=12nm$ \cite{Reyren2009,Ueno2014} we thus assume that on the LAO side of the superconducting layer, defined as $z=0$, the gap $\Delta$ vanishes, since $\Delta$ definitely has to vanish at the interface to the vacuum. On the STO side of the superconducting layer ($z=d$) on the other hand, an interface with a metallic layer can be established. This leads to the boundary condition  $\dro{\Delta^z}z |_{z=d} =0$ \cite{tinkham_intro_to_SC_book}. The geometry is illustrated in figure~\ref{fig:schematic_gap_at_interface}.
From this we may decompose $\Delta^z$ into its Fourier components
\begin{align}
  \label{eq:8}
  \Delta^z (z) &= \sum_{n=1}^\infty \sin\frac{(2n+1)\pi z}{2d} \Delta_{n}^z
  .
\end{align}
Separation of variables then results in different Fourier components for $ f(x,z,\omega) $, for which we obtain
\begin{align}
  \label{eq:f_in_terms_of_delta_with_SOC_fourier_cmtps}
  f_{n}(x,\omega)=\frac{\Delta(x) \Delta_{n}^z}{\omega+\pi H D_m/\phi_0 + \frac18{D} ((2n+1)\pi/d)^2 + 2 D \nu^2}
  .
\end{align}

The new terms $\frac18 D_m\left(\frac{(2n+1)\pi}d\right)^2 $ and $2D\nu^2$ effectively shift the magnetic field by a positive amount and this shift increases for increasing values of $n $.
We may now solve for $H_{c2} $ for each Fourier component $f_{n} $ independently. Since $ \frac18 D_m\left(\frac{(2n+1)\pi}d\right)^2 $ effectively shifts the magnetic field upwards, it is clear that the term with $n = 0 $ will have the largest $H_{c2} $ associated with it. And since we are only interested in the onset of superconductivity, 
we are thus only interested in the most stable component, given by $n = 0 $. 
Re-instating the band index $m$ we obtain for the two bands $m=1,2$ the  following equations
\begin{align}
  f_{m}(x,\omega)=\frac{\Delta_m(x) }{\omega+\pi H D_m/\phi_0 + \frac18{D_m} (\pi/d)^2 + 2 D_m \nu^2}
  .
\end{align}

\begin{figure}[tb]
  \centering
  \includegraphics[width=8cm]{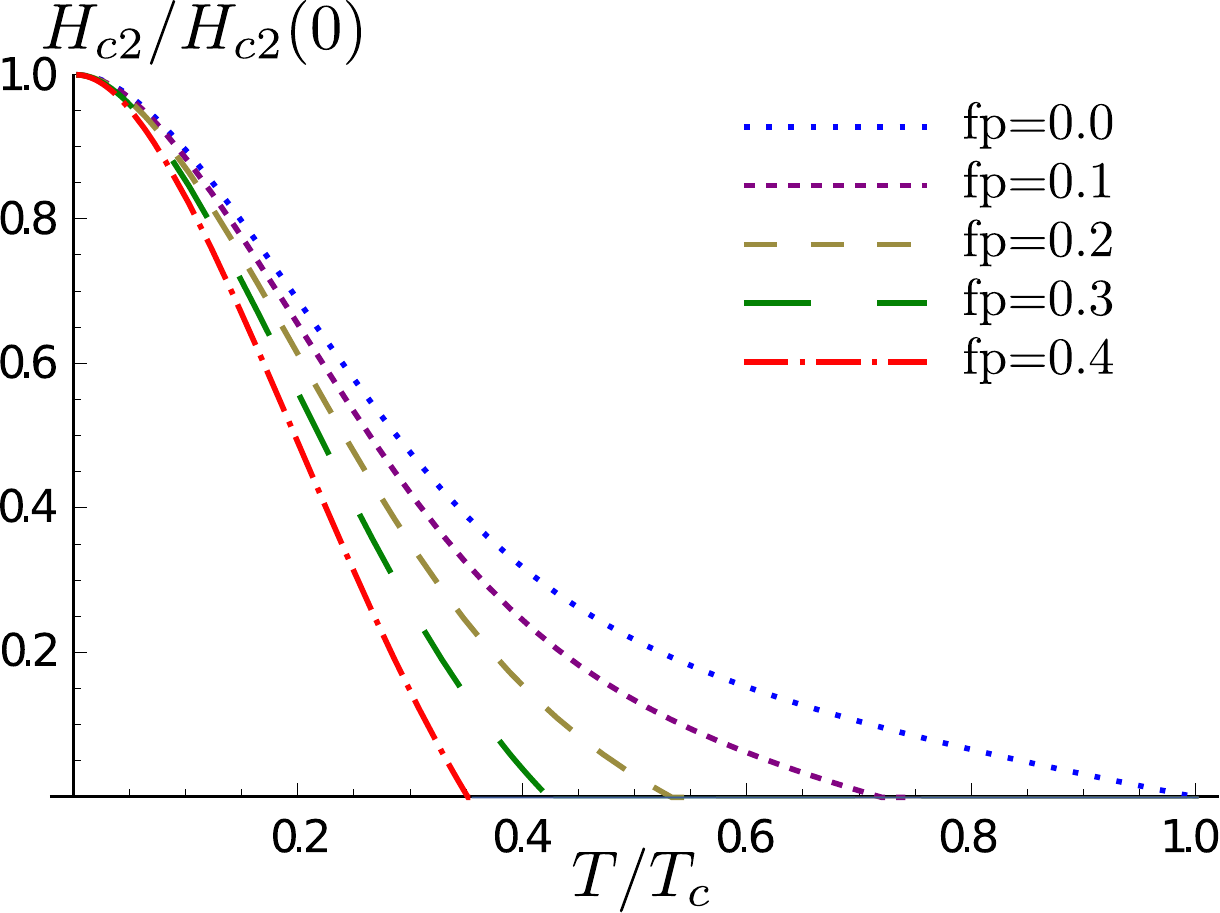}
  \caption{$H_{c2}$ as a function of $T$ for different values of the finite size parameter $fp= \frac{D_1 \pi}{16d^2 T_c}$. Here $\lambda_{11}=0.14, \lambda_{22}=0.13, \lambda_{12}=0.02$ and $\eta=0.05$. As the system becomes increasingly two-dimensional, the critical temperature is reduced. Also, the characteristic low field behaviour disappears, making it difficult to distinguish the single band from the two-band case. $T_c$ refers to the critical temperature for $fp=0$, such that the reduction in the critical temperature due to the finite size becomes apparent.}
  \label{fig:2d_as_fn_of_finite_size}
\end{figure}

\begin{figure}[tb]
  \centering
  \includegraphics[width=8cm]{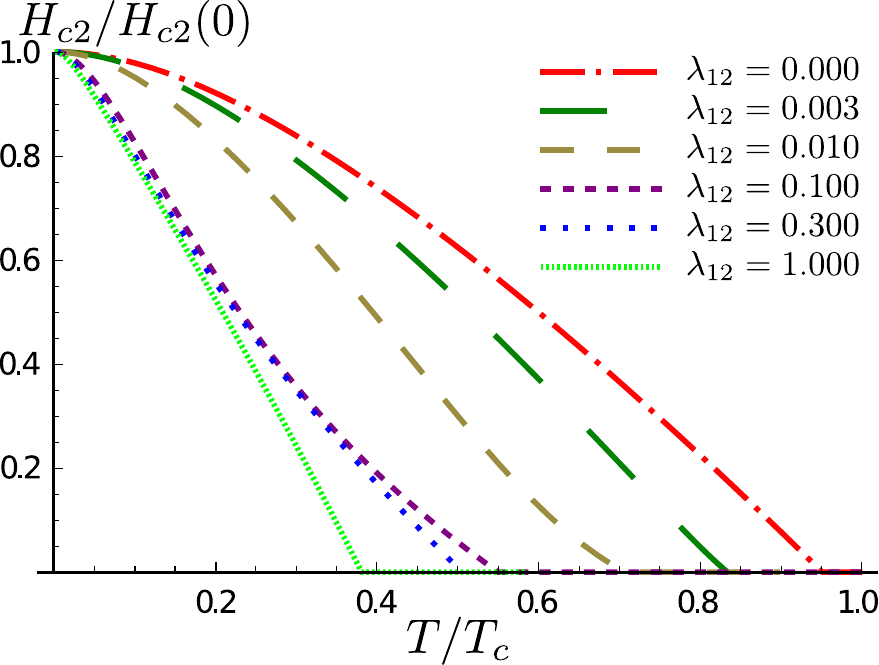}
  
  \caption{$H_{c2}(T)$ for a fixed finite size parameter $fp=0.2$. Otherwise theparameters are the same as in figure~\ref{fig:fix_l11=l22_vary_l12}. As in figure~\ref{fig:2d_as_fn_of_finite_size}, $T_c$ refers to the critical temperature for $fp=0$.}
  \label{fig:hc2_2d_as_fn_of_l12}
\end{figure}

We can now obtain the upper critical field using the formalism described in section~\ref{sec:details-calculation}, with the only difference being that we redefine the quantity $h $ in equation~\eqref{eq:gures_cond_for_Hc2} as
\begin{align}
  h=\frac{H D_1}{2\phi_0 T_c t} + \frac{D_1\pi}{16d^2T_c t}  +  \frac{D_1\nu^2}{\pi T_c t}\label{eq:9}
  .
\end{align}
We define the finite thickness parameter $fp=P_c + P_{soc}$ where $P_c=\pi D_1/(16d^2 T_c) $ encodes the effect of the confining energy and $P_{soc}= D_1\nu^2/(\pi T_c)$ the spin-orbit coupling. Both arise from the fact that the system is inhomogeneous in z-direction.
When the finite thickness parameter $fp$ is appreciable, it can lead to a suppression of the characteristic two-band temperature dependence, as shown in figure \ref{fig:2d_as_fn_of_finite_size}. This is because effectively the low field behaviour (or alternatively high-temperature behaviour) is cut out.
Since experiments indicate that the critical temperature is not much decreased in the interface system as compared to the bulk system, we can assume that the finite thickness parameter is at most on the order unity or smaller.
Figure~\ref{fig:hc2_2d_as_fn_of_l12} shows $H_{c2}(T)$ for a fixed parameter $fp=0.2$ and investigates the shape of the curves for different interband coupling constants $\lambda_{12}$ and otherwise the same parameters as in figure~\ref{fig:fix_l11=l22_vary_l12}. Comparing the blue dotted curves in both figures~\ref{fig:fix_l11=l22_vary_l12} and \ref{fig:hc2_2d_as_fn_of_l12}, we see that in certain cases the finite thickness of the conducting layer can even make the change in curvature more apparent.

The overall shape of the curves, and in particular the qualitative behaviour, is thus the same in the two-dimensional and in the three-dimensional case. A simple estimate of the parameters $P_c$ and $P_{soc}$ gives $P_c=4.3$ and $P_{soc}=1.6$,  $fp$, where $D=\tfrac13 l v_F$ \cite{Usadel1970} was used and the parameters $\xi=70nm$, \cite{Reyren2009}, $l=25nm$ \cite{BenShalom2009}, $v_F=15 km/s$\cite{Lin2013}, $\alpha=3\times 10^{-12} eVm$ \cite{Caviglia2010} and $T_c=0.3K$ \cite{Caviglia2008} were chosen. The resulting values $fp=5.9$ is considerably larger than what we expect from the experimentally only modest decrease in $T_c$, but within the accuracy that might be expected from such a simple estimate. Note, however, that in our estimate the effects of spin-orbit coupling are weaker than those of the finite size corrections

It has been reported that  the superconducting layer at a (110) oriented interface may be considerably thicker than that found at the (001) interface. In those cases it was found that $d\approx 24-30nm$ \cite{Herranz2013}. For $d=30nm$ we find that $P_c=0.68$. Although we cannot trust the quantitative estimates of our finite size parameter, in the case of a (110) oriented interface we expect an $H_{c2}$ curve which is closer to that of bulk STO.


\section{Discussion}
\label{sec:discussion}

Recent experiments by Richter et al \cite{Richter2013} seem to indicate the presence of only one set of coherence peaks in planar tunneling into LAO/STO, at $\Delta_1 \sim 60 \mu eV$. The correct implication hence  was made  that the interface superconductivity is consistent with the single band effect. We point out though, that the expected second superconducting gap is expected to be on the order of $\Delta_2 \sim 25 \mu eV$ and  would be below the observed lifetime broadening on the order of $\Gamma \sim 30-40 \mu eV$.

Experiments by Bert et al \cite{Bert2012} on the superfluid density at the LAO/STO interface have so far ruled out multiband superconductivity with very different gap sizes. 
However, the superfluid density $\rho_s(T) $, is most useful for detecting multiband superconductivity when the coupling constants in the two bands are quite different. This is because for a slow initial growth of $\rho_s(T)  $ to be observed around $T_c $, the characteristic signature, a second gap must open for some $ T<T_c$ \cite{Prozorov2011}. If the two coupling strengths are very similar, the two gaps will open at roughly the same temperature and a signature of multiband superconductivity is hard to detect.
Since the upper critical field is most sensitive to multiband superconductivity when the coupling constants in the two band are very similar (see section~\ref{sec:more-gener-param}), the superfluid density and the upper critical field are thus complementary probes for multiband superconductivity which work in opposite regimes.

We therefore suggest that this proposed study of $H_{c2}(T)$   would be a useful alternative probe to detect multiband superconductivity. 

\section{Conclusion}
\label{sec:conclusion}

In this paper we investigated the temperature dependence of the upper critical field in two-band superconductors, with a view to finding an experimental criterion for the presence of two-band superconductivity. We have found that, in particular in the disordered regime, $H_{c2} (T) $ exhibits a characteristic behaviour which is qualitatively different from that of single band superconductors. Experiments may thus be able to use this property to confirm that STO is indeed a two-band superconductor. This tool is particularly useful for the investigation of the superconductivity at the interface between LAO and STO as it will help to relate it to the  superconductivity in bulk STO.

\section{Acknowledgements}
We are grateful to K. Behnia, R. Fernandes, J. Haraldsen, J.X. Zhu and S. Lederer for useful discussions and H. Haraldsen, K. Moler and K. Behnia for comments on the draft. We would also like to thank K. Behnia for showing us some of the data in Ref.\cite{Lin2014} prior to publication. Work was supported by Nordita, VR 621-2012-2983 and ERC 321031-DM. Work  at Los Alamos was supported by the Office of Basic Energy Sciences and by LDRD.

\appendix

\section{Induced triplet superconductivity}
\label{sec:induc-tripl-superc}

We now look at the triplet component of superconductivity that is induced by spin-orbit coupling and show that it is not relevant for our calculations. We treat spin-orbit coupling in a perturbative way and assume that it is smaller than the Fermi energy. The self consistent expression for the gap $\Delta $ within a single band in the the absence of spin-orbit coupling is given by \cite{altland_simons_cmft_book}
\begin{align}
  \Delta= V \frac{T}{L^d}\sum_{p,n} \frac{\Delta}{\omega_n^2 + \xi_p^2 + \Delta^2}
\end{align}
with $\xi_p=\frac{p^2}{2m}-\mu$ and $V$ is the interaction potential . We now take spin-orbit coupling into account. We thus write $ \Delta$ as a matrix according to equation~(\ref{eq:delta_sing_trip}) and $ \xi$ turns into
\begin{align}
  \xi\to \xi + \alpha (\bm{k} \times \bm {\sigma})=\xi + \alpha(k_x \sigma_y - k_y \sigma_x)
\end{align}
If we assume the existence of a singlet gap $ \Delta_s$ we can obtain to lowest order in $\alpha $ the perturbed expression for $\tilde\Delta $
\begin{align}
  \tilde\Delta = V \frac T{L^d} \sum_{p,n}\frac{i\sigma_y \Delta_s (\omega_n^2 + \xi_p^2 + \Delta^2) - 2i\alpha k_x \xi_p \Delta_s \sigma_0 + 2\alpha k_y \xi_p \Delta_s \sigma_z }{ (\omega_n^2 + \xi_p^2 + \Delta^2) ^2}
\end{align}
The induced triplet components are thus given by
\begin{align}
  \Delta^a&=  V \frac T{L^d} \sum_{p,n} \frac{2\alpha k_y \xi_p \Delta_s}{(\omega_n^2 + \xi_p^2 + \Delta^2) ^2}\\
  \Delta^b&=  V \frac T{L^d} \sum_{p,n} \frac{2\alpha k_x \xi_p \Delta_s}{(\omega_n^2 + \xi_p^2 + \Delta^2) ^2}\\
  \Delta^c&=0
  .
\end{align}
Since $\Delta^a$ and $\Delta^b$ contain a sum over all $k_y $ or $k_x$ values, they vanish. This leads us to the conclusion that also the triplet pairing amplitudes $f^a,f^b$ and $f^c$ vanish, in the approximation that interaction $V$ has no p wave components. If there are small p wave components the induced triplet components will be small in proportion. 

\section*{References}

%\bibliography{library}

\end{document}